# Role of Loading Device on Single-Molecule Mechanical Manipulation of Free Energy Landscape

Gwonchan Yoon, Sungsoo Na, and Kilho Eom[*]

*Department of Mechanical Engineering, Korea University, Seoul 136-701, Republic of Korea*

**ABSTRACT**

Single-molecule mechanical manipulation has enabled the quantitative understanding of the kinetics of bond ruptures as well as protein unfolding mechanism. Single-molecule experiments with theoretical models have allowed one to gain insight into free energy landscape for chemical bond and/or protein folding. For mechanically induced bond rupture, the bond-rupture kinetics may be governed by loading device. However, the role of loading device on the kinetics of mechanical rupture has been rarely received much attention until recently. In this work, we have theoretically and/or computationally studied the effect of loading-device stiffness on the kinetics of mechanical unfolding. Specifically, we have considered a one-dimensional model for a bond rupture whose kinetics is depicted by Kramers' theory. It is elucidated that the kinetics of bond rupture is determined by force constant of loading device. The Brownian dynamics simulation of a bond rupture is considered in order to validate our theory. It is illustrated that the mean rupture force is dependent on the force constant of a loading device, such that increase in the loading-device stiffness leads to the higher bond rupture force. Moreover, we have taken into account the computational simulation of mechanical unfolding of a small protein, i.e. *β*-hairpin. Our simulation shows that unfolding force is highly correlated with the stiffness of a loading device. Our numerical and theoretical studies highlight the significance of a loading device on the kinetics of mechanical unfolding of a chemical bond and/or a folded domain in the single-molecule mechanical experiments.

**Introduction**

Single-molecule mechanical manipulation has been conceived as one of important nanomechanical tools that allow the characterization of biochemical reactions (1-6). Over a last decade, single-molecule mechanical manipulation using a loading device such as atomic force microscopy [AFM] (7-9) and/or optical tweezer [OT] (10-12) has provided the mechanisms of bond rupture and/or protein unfolding, which plays a critical role on the biological function. The principle of single-molecule mechanical manipulation is that a single-molecule exhibiting a chemical bond or a folded domain is mechanically stretched via a loading device. In general, the mechanical stretching process is usually under regime of non-equilibrium process (13), so that single-molecule mechanical manipulation does not provide the direct insight into free energy landscape relevant to a chemical bond or a folded domain.

There have been current attempts (3-6) to extract the free energy landscape from the single-molecule mechanical manipulation based on theoretical models. For instance, Bell (14) has introduced the simple, theoretical model of a bond rupture with assuming that bond rupture is well depicted by diffusion-like process, i.e. $k(F) = k_0\exp(F/F_c)$, where $k(F)$ is the rate constant for a bond rupture as a function of mechanical force $F$, $k_0$ is the rate constant at zero force, and $F_c$ is defined as $F_c = k_BT/\Delta x$ with $k_B$, $T$, and $\Delta x$ being the Boltzmann's constant, the absolute temperature, and the energy barrier width at zero force, respectively. Bell's model (14) has been successfully employed to characterize the free energy landscape from single-molecule mechanical experiments (4, 5). However, Bell's model is only appropriate for a case, where a molecule is slowly stretched. In other words, if a molecular is pulled fast enough that stretching processes quickly reaches the non-equilibrium process, Bell's model is not suitable to describe the bond rupture mechanism. In a recent decade, there have been theoretical approaches that are able to explain the non-equilibrium process of bond rupture. For instance, Evans and Ritchie (15) had developed the theoretical framework, where Kramers' theory (16) was used to describe the bond rupture mechanism. Hummer and Szabo (17) provided the theoretical framework that enables the interpretation of non-equilibrium protein unfolding mechanics. In the similar spirit, Dudko *et al.* (18) had employed the Garg's argument (19) on escape field theory in order to understand the bond rupture mechanism under non-equilibrium process. Moreover, with presumed free energy landscape, Dudko *et al.* (20) have theoretically obtained the rate constant for mechanical bond rupture under both nearly equilibrium and fully non-equilibrium processes. Further, theoretical models relevant to non-equilibrium process, such as Jarzynski's inequality (21) and/or Crooks' theorem (22), have been considered to extract the free energy landscape for a chemical bond or a protein folding from single-molecule mechanical experiments (11, 23).

---

[*] The author to whom the correspondence should be addressed. E-mail: kilhoeom@korea.ac.kr or kilhoeom@gmail.com

Extraction of free energy landscape from single-molecule experiments has been considered using theoretical models as described above. Here, theoretical models have been developed with assumption of soft loading devices (20, 24, 25), since loading device used in single-molecule mechanical manipulation exhibits the force constant of < 0.1 N/m [Ref. (6, 26)]. In general, mechanical behavior of a single-molecule depends on the loading devices. For example, AFM possesses the force constant of < 0.1 N/m and the available loading rate for AFM is usually in the range of $10^5$ pN/s to $10^6$ pN/s. On the other hand, the force constant of OT is much less than that of AFM by three orders, and the accessible loading rate is ~1 pN/s, much less than that available for AFM. This implies that single-molecule stretching by OT is almost the equilibrium process (27), whereas AFM-driven single-molecule manipulation is under non-equilibrium process. This indicates that extraction of free energy landscape from single-molecule mechanical manipulation has to be carefully taken into account by considering the effect of loading device.

Despite a lot of efforts in theoretical descriptions on bond rupture mechanism relevant to protein unfolding mechanics, there has been rarely explored to study the effect of loading device on the kinetics of bond rupture and/or protein unfolding mechanics. Most theoretical works (17, 18, 20, 25) have neglected the effect of loading device by assuming that loading device is very soft. Specifically, the theoretical works (17, 18, 20, 25, 28) assumes that a chemical bond (or equivalently a folded protein domain) is mechanically stretched by a very soft loading device. For such an assumption, the kinetic rate for bond rupture becomes independent of loading device. This can be clearly demonstrated using a simple harmonic model for a bond rupture. In this case, the effective potential energy for a mechanically stretched chemical bond is given by $V_{eff} = (1/2)s_m x^2 + (1/2)s_L(x - \lambda)^2$, where $s_m$, $s_L$, $x$, and $\lambda$ indicates the force constant of a chemical bond (or a folded domain), the force constant of a loading device, a reaction coordinate, and a control parameter defined as $\lambda = vt$ with $v$ being pulling rate, respectively. For soft loading device and weak force, the energy barrier becomes $\Delta V \approx -(\partial V/\partial \lambda)\Delta\lambda = -F\Delta\lambda$, where $F = s_{eff}\lambda$ with $s_{eff}$ being an effective force constant given as $s_{eff} = (1/s_m + 1/s_L)^{-1}$, and $\Delta\lambda$ represents the difference in $\lambda$ between bonded and denatured states. Here, $\Delta\lambda$ can be treated as a constant. Consequently, the rate constant for bond rupture under nearly equilibrium process is given from Bell's theory such as $k(t) = k_0 \exp\left(s_{eff} v t \Delta\lambda / k_B T\right) = k_0 \exp\left(\dot{F} t \Delta\lambda / k_B T\right)$, where $\dot{F}$ is the loading rate given by $\dot{F} = s_{eff} v$. This shows that such approximation (i.e. soft loading device and weak force) is unable to capture the effect of loading device on the kinetic rate of bond rupture.

To our best knowledge, the role of loading device on bond rupture and/or protein unfolding mechanism has not been studied except recent few theoretical studies (24, 29). In these studies (24, 29), a simple theoretical models based on harmonic potential for a bond [e.g. Ref. (29)] and/or a phenomenological model of bond rupture [e.g. Ref. (24)] have been considered to gain insight into the role of loading device on the bond rupture mechanism. However, these theoretical works have not been validated using single-molecule experiments/simulations on bond ruptures and/or protein unfolding mechanics.

In this work, we have studied the kinetics of mechanical unfolding of a chemical bond or a folded protein domain using physical model based on. Kramers' theory (16) and computational simulations such as Brownian dynamics simulation of bond rupture and/or protein unfolding mechanics. Our theoretical model provides that bond rupture mechanism usually described by kinetic rate, $k$, probability distribution of bond rupture force, $p(F)$, and mean rupture force, $<F>$, is governed by control parameters of loading device, e.g. force constant and loading rate. Further, our Brownian dynamics simulation of bond rupture and/or protein unfolding mechanics illustrates that mechanical unfolding mechanism is determined by control parameters for loading device. This is attributed to the principle that perturbation of free energy landscape due to the bond rupture (or equivalently, protein unfolding) is dominated by potential energy profile of loading device. Our study highlights the significant role of loading device on the bond rupture mechanism and/or protein unfolding mechanism. This implies that single-molecule mechanical manipulation has to be carefully interpreted by consideration of loading device effect.

**Bond Rupture Model: One-Dimensional Model**

In order to understand the bond rupture mechanism, we would like to consider the simple, theoretical model that describes a chemical bond whose free energy landscape is represented with respect to reaction coordinate $x$. Here, free energy landscape for a chemical bond is depicted as $U_0(x)$, and the potential energy for a loading device to perturb the free energy landscape leading to bond rupture is given by $V_L(x) = (s_L/2)(x - \lambda)^2$, where $\lambda$ is a control parameter, for a loading device, defined as $\lambda = vt$ with $v$ being a pulling rate and $t$ being the time. The probability density to have an intact chemical bond is well described by Smolouchowski equation (30).

$$\frac{\partial \psi}{\partial t} = \frac{\partial}{\partial x} D \left[ k_B T \frac{\partial \psi}{\partial x} + \psi \frac{\partial V_{eff}}{\partial x} \right] \equiv \wp \psi \qquad (1)$$

where $\psi(x, t; x_0)$ is the probability density to possess an intact bond at time $t$ and reaction coordinate $x$ under the initial position of $x_0$, $D$ is the diffusion coefficient, $k_B T$ is the thermal energy, $V_{eff}(x)$ is an effective potential field for a mechanically stretched molecule such as $V_{eff}(x) = U_0(x) + V_L(x)$, and a symbol $\wp$ indicates the differential operator defined as $\wp = (\partial/\partial x)[D\{k_B T(\partial/\partial x)+ (\partial V_{eff}/\partial x)\}]$. The probability for an intact bond at time $t$, i.e. $Q(t; x_0)$, is given by

$$Q(t; x_0) = \int dx \cdot \psi(x, t; x_0) = \int dx \cdot \exp[t\wp] \cdot \delta(x - x_0) \qquad (2)$$

where $\delta(x)$ is the Dirac delta function. Consequently, the probability distribution for a bond rupture time, $p(t; x_0)$, is represented as $p(t; x_0) = -dQ(t; x_0)/dt$, and then the mean unfolding time, $\tau(x_0)$, is given by

$$\tau(x_0) = \int_0^\infty dt \cdot t \cdot p(t; x_0) = \int_0^\infty dt \int dx \cdot \delta(x - x_0) \cdot \exp[t\wp^\dagger(x)] \qquad (3)$$

Here, the integration of parts was used, and a symbol $\wp^\dagger$ represents an adjoint to a differential operator $\wp$. The rate constant for a bond rupture, $k$, is thus obtained as

$$k(t) \cong \frac{D\omega_b(t)\omega_{ts}(t)}{2\pi k_B T} \exp\left[-\frac{\Delta V(x; t)}{k_B T}\right] \qquad (4)$$

This formula is renowned as Kramers' theory (16) that was revisited by Evans and Ritchie (15). Here, $\omega_b(t)$ and $\omega_{ts}(t)$ represent the natural frequencies at an equilibrium for bonded state and a local equilibrium at which barrier crossing (that is, bond rupture) occurs, and $\Delta V(x; t)$ is the energy barrier defined as $\Delta V(x; t) = V_{eff}(x_{ts}; t) - V_{eff}(x_b; t)$, where $x_b$ and $x_{ts}$ indicate the reaction coordinates at an equilibrium for bonded state and a local equilibrium at which bond rupture happens, respectively. It should be recognized that Kramers' theory (16) is a generic model to describe the bond rupture over the entire regime of pulling rates. In general, Bell's model (14) only captures the bond rupture event when a mechanical stretching of a chemical bond is almost equilibrium process (i.e. very slow pulling rate), while Garg's escape field theory (19) is only appropriate to describe the bond rupture when a chemical bond is stretched by a critical loading rate at which energy barrier instantaneously disappears (18).

**Theoretical Analysis on Bond Rupture: Linear-Cubic Potential**

We consider the one-dimensional free energy landscape for a chemical bond (or equivalent to folded protein domain) represented in the form of $U_0(x) = (3/2)\Delta G_0(x/\Delta x) - 2\Delta G_0(x/\Delta x)^3$, where $\Delta G_0$ is an energy barrier at zero force, and $\Delta x = x_{ts}^0 - x_b^0$, where $x_{ts}^0$ and $x_b^0$ indicate the reaction coordinates at equilibrium for bonded (folded) state and at transition state (at which bond rupture occurs) with zero force, respectively. Here, it should be noticed that free energy landscape $U_0(x)$ is able to capture the free energy landscape for a folded protein domain with respect to reaction coordinate (20). In order to introduce the effective force constant $s_{eff}$ (as shown above), the free energy $U_0(x)$ has to be simplified using harmonic approximations: $U_0(x) \approx (1/2)s_m x^2$, where $s_m = 6\Delta G_0/(\Delta x)^2$. Subsequently, the effective force constant is given as $s_{eff} = (1/s_m + 1/s_L)^{-1} = 6\Delta G_0 s_L/[6s_L(\Delta x)^2 + 6\Delta G_0]$. As the free energy landscape begins to be perturbed by a loading device, the reaction coordinates $x_{ts}$ and $x_b$ becomes dependent on the control parameters of loading device. One can straightforwardly find the reaction coordinates $x_{ts}(\lambda)$ and $x_b(\lambda)$ from $\partial V_{eff}/\partial x = 0$.

$$\bar{x}_b(\bar{\lambda}) = \varepsilon - 0.5\sqrt{1 - 8\varepsilon\bar{\lambda} + 16\varepsilon^2} \qquad (5.a)$$

$$\bar{x}_{ts}(\bar{\lambda}) = \varepsilon + 0.5\sqrt{1 - 8\varepsilon\bar{\lambda} + 16\varepsilon^2} \qquad (5.b)$$

Here, we have introduced the dimensionless parameters using normalizations such as $\bar{x}_b \equiv x_b/\Delta x$, $\bar{x}_{ts} \equiv x_{ts}/\Delta x$, $\bar{\lambda} = \lambda/\Delta x \equiv vt/\Delta x$, and $\varepsilon = s_L(\Delta x)^2/12\Delta G_0 \equiv s_L/s_m$. Herein, $\bar{\lambda}$ is a dimensionless pulling rate, and $\varepsilon$ is a non-dimensional force constant of loading device. For the slow pulling of a bond with very soft spring, the reaction coordinates $x_{ts}(\lambda)$ and $x_b(\lambda)$ asymptotically approaches to $x_{ts} \approx \varepsilon\Delta x - 0.5\Delta x$ and $x_b \approx \varepsilon\Delta x + 0.5\Delta x$. One can easily discover that, for slow pulling with very soft loading devices, the energy barrier width is independent of control parameters of a loading device and given as $x_{ts} - x_b \approx \Delta x$. This indicates that Bell's model using the constant energy barrier width is the limiting case to the slow pulling with soft loading device. With reaction coordinates given by Eq. (5), the energy barrier under the mechanical stretching induced by a loading device is represented in the form of

$$\Delta \bar{V} \equiv \frac{\Delta V}{\Delta G_0} = \frac{V_{eff}(x_{ts};t) - V_{eff}(x_b;t)}{\Delta G_0} = \left(1 - 8\varepsilon\bar{\lambda} + 4\varepsilon^2\right)^{3/2}$$
$$\equiv \left[1 - 8(1+\varepsilon)\dot{F}\tau + 4\varepsilon^2\right]^{3/2} \qquad (6)$$

where $\dot{F}$ is the loading rate defined as $\dot{F} = s_{eff}v$, and $\tau$ is the normalized time-scale defined as $\tau = t/s_m\Delta x$. For slow pulling with soft loading device, it is shown that energy barrier asymptotically approaches to $\Delta V \approx \Delta G_0(1 - 12\varepsilon\bar{\lambda}) + O(\bar{\lambda}^2) = \Delta G_0 - s_L\lambda\Delta x \equiv \Delta G_0 - (1+\varepsilon)(\Delta x)\dot{F}t$. For the limiting case of soft loading device, i.e. $\varepsilon \ll 1$, the energy barrier becomes $\Delta V \approx \Delta G_0 - (\Delta x)\dot{F}t$, which is consistent with Bell's model on bond rupture. This indicates that the limiting case is unable to capture the effect of loading device on the kinetics of bond rupture. With Kramers' theory dictated by Eq. (4), the kinetic rate for bond rupture driven by loading device is given by

$$\bar{k}(\dot{F},\varepsilon) = \bar{T}^{-1}\sqrt{4\varepsilon^2 - 8\varepsilon\bar{\lambda} + 1} \exp\left[-\frac{\pi}{3\bar{T}}\left(1 - 8\varepsilon\bar{\lambda} + 4\varepsilon^2\right)^{3/2}\right]$$
$$= \bar{T}^{-1}\sqrt{4\varepsilon^2 - 8(1+\varepsilon)\dot{F}\tau + 1} \exp\left[-\frac{\pi}{3\bar{T}}\left\{1 - 8(1+\varepsilon)\dot{F}\tau + 4\varepsilon^2\right\}^{3/2}\right] \qquad (7)$$

where $\bar{k}(\dot{F},\varepsilon)$ is the normalized rate constant for bond rupture, defined as $\bar{k} = k(t) \cdot (\Delta x)^2/D$, and $\bar{T}$ is the dimensionless temperature defined as $\bar{T} = \pi k_B T/3\Delta G_0$. As described in Eq. (7), the kinetic rate for bond rupture is determined by two significant parameters $\dot{F}$ (i.e. loading rate) and $\varepsilon$ (i.e. stiffness of loading device). In case of soft loading device (i.e. $\varepsilon \ll 1$) and slow pulling rate (i.e.

$\dot{F}\tau \ll 1$), the kinetic rate for bond rupture approaches to $k(t) \approx \overline{T}^{-1} \exp\left[-\left(\pi/3\overline{T}\right)\left(\Delta G_0 - \dot{F}t\Delta x\right)\right]$. This indicates that, with approximations of soft loading device and slow pulling, the kinetic rate for bond rupture becomes independent of the force constant of a loading device.

Once the kinetic rate is explicitly obtained, we need to know the probability distribution of bond rupture force, since single-molecule experiments provide such probability distribution and its corresponding mean value of bond rupture force. With assumption that bond rupture is described as first-order phase transition, we can find the probability to have an intact bond at time $t$, $Q(t)$, from a relation of

$$\frac{dQ}{dt} = -k(t) \cdot Q(t) \tag{8}$$

It should be noticed that the force exerted by loading device at time $t$ is given as $F(t) = s_L vt$. The probability $Q(F)$ is represented in the form of

$$Q(F) = \exp\left[-\frac{1}{s_L v}\int_0^F k(F')dF'\right] \tag{9}$$

Here, the rate constant $k$ and the probability $Q$ are expressed with respect to force $F(t)$ rather than time $t$. The probability distribution of bond rupture forces is given by

$$p(F) = -\frac{k(F)}{s_L v} \exp\left[-\frac{1}{s_L v}\int_0^F k(F')dF'\right] \tag{10}$$

The mean rupture force is, thus, calculated from a relation of

$$\langle F \rangle = \int_0^\infty F \cdot p(F) dF = \int_0^\infty -\frac{F \cdot k(F)}{s_L v} \exp\left[-\frac{1}{s_L v}\int_0^F k(F')dF'\right] dF \tag{11}$$

The mean rupture force depicted by Eq. (10) is well defined regardless of loading-device stiffness. For soft loading device, the asymptotic expression of mean rupture force is given in Ref. (25).

**Theoretical Calculations: Bond Rupture Kinetics and Rupture Forces**

Fig. 1a shows the free energy landscape (described by linear-cubic potential) that is perturbed by mechanical stretching using loading device with its different stiffness. It is shown that free energy landscape for mechanically stretched bond is governed by stiffness of loading device such that stiffer loading device induces the more perturbation in the free energy landscape. This implies that stiffer loading device drives the fast kinetics of bond rupture. The inset of Fig. 1a depicts the normalized energy barrier, $\Delta \overline{V}$, as a function of dimensionless force constant of loading device, $\varepsilon$, and normalized pulling rate $\overline{\lambda}$. It is illustrated that energy barrier is decreasing when the stiffness of loading device increases. In other words, stiffer loading device triggers the faster bond rupture, indicating that energy barrier disappears fast when stiff loading device is employed for mechanical stretching. As shown Fig. 1b, we have theoretically suggested that the kinetic rate for bond rupture is strongly affected by force constant of loading device. This indicates that mechanical stretching of a bond undergoes the non-equilibrium process when stiff loading device is used. With model parameters (for details, see caption of Fig. 1), theoretical calculation based on Eqs. (10) and (11) provides the probability distribution of bond rupture forces (shown in Fig. 1c) and the mean rupture forces (shown in Fig. 1d) as a function of loading rate, $\dot{F}$, and force constant of loading device, $s_L$. Our theoretical model vividly demonstrates that the effect of loading device has to be carefully examined when single-molecule experimental data (e.g. bond rupture force, probability distribution of rupture force, etc.) is interpreted to extract the free energy landscape.

**Brownian Dynamics Simulations on Bond Ruptures: Loading-Device Stiffness vs. Rupture Forces**

For an insight into the generic bond rupture mechanism, we have employed the Brownian dynamics simulation of bond rupture. It should be noticed that our theoretical calculations assume that free energy landscape is in the form of linear-cubic potential. However, in general, the free energy landscape for a chemical bond is not linear-cubic form but depends on the physical model of a bond. Moreover, in our simulation, it should be recognized that Brownian dynamic simulation is valid for a case, where energy dissipation due to friction between environment and a chemical bond dominates the reaction process. That is, the inertia effect is assumed to be insignificant in the bond rupture process. Our simulation also presumes that the free energy landscape for a chemical bond is depicted by Morse potential, which is suitable to describe the chemical bond in biological system, e.g. hydrogen bond connecting two phosphates in double-stranded DNA (31). Fig. 2a presents the force $F(t)$, which is exerted by a loading device, with respect to time $t$ for a mechanically stretched chemical bond with a loading rate of $\dot{F} = 1$ nN/s. Here, the loading rate is given by a relation of $\dot{F} = s_{eff} v$, where $s_{eff}$ is an effective force constant, which is obtained as a slope of the force-extension curve (i.e. $s_{eff} = \partial F/\partial u$ with $F$ and $u$ being the force and the displacement, respectively, i.e. $u = vt$), and $v$ is the pulling rate. For mechanical stretching using stiff loading device, the rebinding of a bond is observed, whereas such rebinding is not found when a chemical bond is stretched by a soft loading device. It is attributed to the fact that stiff loading device induces the smaller energy barrier $\Delta V$ (e.g. Fig. 1a), which allows a chemical bond to undergo the phase transition easily via thermal fluctuation. However, energy barrier crossing by soft loading device is not as fast as that using stiff loading device, which may result in the irreversible bond rupture driven by soft loading device. As a consequence, the rebinding of denatured bond is not found when soft loading device is used for mechanical stretching. Fig. 2b shows the probability distributions of bond rupture forces with respect to control parameters of loading device, i.e. loading rate and force constant. The mean rupture forces as a function of loading rate and force constant of loading device are depicted in Fig. 2c. It is found that mean rupture force is increased when stiffer loading device is utilized. This suggests that the force constant of loading device has to be importantly considered when one attempts to extract the parameters for free energy landscape, e.g. energy barrier and/or energy barrier width, based on single-molecule mechanical stretching experiments. Our simulation of bond rupture provides that bond rupture mechanism is determined by control parameters of loading device, consistent with our theoretical model.

**Protein Unfolding Mechanics: Unfolding Force Is Governed by Loading-Device Stiffness**

For more realistic situation of protein unfolding mechanics, we have employed the coarse-grained molecular dynamics simulation through which a protein domain is pulled with constant loading rate. For computational efficacy, we have assumed that protein unfolding mechanics undergoes the Brownian dynamics (for details, see Methods). Here, we have considered β-hairpin as a model protein, since it is a small protein to possess the hydrogen bonds that can be ruptured by single-molecule mechanical stretching. Fig. 3a provides the representative curves for force vs. time for mechanical unfolding of β-hairpin. Furthermore, the snapshots in Fig. 3a shows conformations for β-hairpin that is mechanically stretched with loading rate of $\dot{F} = 3 \times 10^{-3}$ N/s and different force constants (i.e. 0.01 N/m and 0.05 N/m) of loading device. Snapshots in Fig. 3a provide that unfolding pathway is unaffected by force constant of loading device, which implies that different force-extension curves (in Fig. 3a) is not attributed to the unfolding pathway. That is, the unfolding pathway is purely determined by geometry of hydrogen bonds (32) but not by the loading device effect. It is interestingly found that force-extension relation is strongly dependent on force constant of loading device such that a force drop due to bond rupture is increasing when stiffer loading device is used, whereas force drop due to bond rupture is very small when a soft loading device is employed (see Fig. 3a). This is ascribed to the hypothesis that mechanical stretching using very soft loading device is nearly equilibrium process. For protein unfolding using stiff loading device (i.e. force constant of 0.05 N/m), the significant force drops implies the noticeable energy dissipation, indicating that mechanical stretching using stiff device is under a non-equilibrium process. Moreover, Fig. 3b shows the unfolding force with respect to loading rate and force constant. Here, the unfolding force is measured as the peak force, at which a folded domain becomes to be fully denatured. To our best knowledge, the relationship between unfolding force (for protein unfolding) and force constant of loading device has been rarely studied. It is found that loading device affects the unfolding force such that stiffer loading device increases the unfolding force. This is consistent with our theoretical anticipation based on simple two-state kinetic model using Kramers' theory with simplified free energy landscape, albeit free energy landscape of a protein is generally insufficient to be described by two-state model but it is of complexity, e.g. more than two-state kinetics (12) and ruggedness (33). Our computational simulation results illustrates that protein unfolding mechanics can be manipulated by controlling the force constant of loading device.

**Conclusion**

In conclusion, we have elucidated the role of loading device on the bond rupture mechanism as well as protein unfolding mechanics based on theoretical model and/or computational simulation. We have theoretically provided the exact form of kinetic rate for bond rupture as a function of two controlling parameters – loading rate and force constant. It is shown that rate constant for bond rupture is determined by such two parameters. This indicates that the mechanical response of mechanical bond-rupture from single-molecule pulling experiments has to be interpreted by considering the effect of loading device (i.e. force constant). Until recently, such effect has not been taken into account for interpretation of single-molecule experiments. Moreover, our computational simulations on bond rupture as well as protein unfolding mechanics illustrates that kinetic rate for bond rupture and/or protein unfolding is dominated by two controlling parameters. This is attributed to the fact that mechanical stretching using soft loading device is nearly equilibrium process, whereas mechanical extension by stiff loading device undergoes a non-equilibrium process. In our theoretical work, it should be reminded that our model is very restrictive for interpretation of protein unfolding mechanics because of our assumptions such as one-dimensional, two-state kinetic model, albeit our theoretical model is able to qualitatively describe the protein unfolding mechanism. In order to quantitatively interpret the protein unfolding mechanics, the more realistic free energy landscape model (e.g. multi-dimensionality (34), more than two-state kinetics (12), etc.) will be considered with Kramers' theory for our future work.

**Methods**

***Brownian Dynamics Simulation of Bond Rupture.*** For simulation of bond rupture, we have employed the Brownian dynamics, which discards the inertia effect in the dynamics. The reaction coordinate $x(t)$ under the mechanical stretching is described as

$$x(t + \Delta t) = x(t) + \frac{\Delta t}{\gamma}\left[-\frac{\partial V_{eff}(x)}{\partial x} + \xi(\Delta t)\right]$$

where $\Delta t$ is the time step for numerical integration (here, we set $\Delta t$ = 100 ps), $\gamma$ is the friction coefficient arising from friction between a bond and environment, $V_{eff}$ is an effective potential given by $V_{eff}(x) = U_0(x) + V_L(x)$ with $U_0(x)$ and $V_L(x)$ being the potential fields for a bond and a loading device, respectively, and $\xi(\Delta t)$ is the Gaussian random force driven by thermal fluctuation. It should be noticed that the time step of 100 ps is acceptable based on stability criterion. Herein, we have used the Morse potential to describe a chemical bond, i.e. $U_0(x) = U_0[\{1 - \exp(-2b(x - R_c)/R_c)\}^2 - 1]$, and we set $U_0$ = 0.12 nN·nm, $b$ = 1.5, $R_c$ = 0.24 nm, $\gamma$ = 7.7 × $10^{-6}$ kg/s, $T$ = 293 K.

***Brownian Dynamics Simulation of Protein Unfolding Mechanics.*** The governing equation for protein unfolding mechanics is spiritually identical to that for bond rupture simulation.

$$\mathbf{r}_i(t + \Delta t) = \mathbf{r}_i(t) + \frac{\Delta t}{\gamma}\left[-\frac{\partial V_{eff}(\mathbf{r})}{\partial \mathbf{r}_i} + \mathbf{R}_i(\Delta t)\right]$$

Here, $\mathbf{r}_i$ is the position vector of $i$-th alpha carbon atom, $\mathbf{r}$ indicates the coordinates of all residues, i.e.

$\mathbf{r} = [\mathbf{r}_1, \ldots, \mathbf{r}_N]$, where $N$ is the total number of residues, $V_{eff}(\mathbf{r})$ is an effective potential defined as $V_{eff} = U_0 + V_L$ with $U_0$ and $V_L$ being the potential energies for a protein and a loading device, respectively, and $\mathbf{R}_i(t)$ is the Gaussian random force acting on residue $i$. We have used the time step of $\Delta t = 3$ fs and the pulling rate in the range of $10^7$ nm/s to $10^8$ nm/s. Brownian dynamics description is acceptable for such pulling rate regime (35). We have employed the Gō potential to describe the protein structure.

$$U_0(\mathbf{r}) = \sum_{i=1}^{N-1} \sum_{j=i+1}^{N} \left[ \left\{ \frac{k_1}{2}(d_{ij} - d_{ij}^0)^2 + \frac{k_2}{4}(d_{ij} - d_{ij}^0)^4 \right\} \delta_{j,i+1} + 4e_0 \left\{ \left(\frac{\sigma}{d_{ij}}\right)^{12} - \left(\frac{\sigma}{d_{ij}}\right)^6 \right\} (1 - \delta_{j,i+1}) \right]$$

where $d_{ij}$ is the distance between two residues $i$ and $j$, $k_1$ and $k_2$ are force constants for harmonic and quartic potentials, respectively, for a covalent bond, $e_0$ is the energy parameter for native contact, and $\sigma$ is a length scale for native contact, $\delta_{ij}$ is the Kronecker delta, and superscript 0 indicates the equilibrium state. Herein, we have set $d_{ij}^0 = 3.8$ Å, $k_1 = e_0/\text{Å}^2$, $k_2 = 100 e_0/\text{Å}^2$, and $e_0 = 68$ pN·Å [Ref. (36)]. It should be noted that, before simulation of mechanical stretching, we have used the energy minimization to find the equilibrium conformation. For simulation of mechanical stretching of a protein, we have prescribed the potential energy for a loading device as $V_L = (s_L/2)(vt - x)^2$, where $x$ is the distance between N- and C-termini, i.e. $x = |\mathbf{r}_1 - \mathbf{r}_N|$. The mechanical stretching of a protein is simulated from equation of motion, and then force-extension curve is obtained. The snapshots of conformations were generated using Visual Molecular Dynamics. It should be noted that, in our simulation of protein unfolding, the β-hairpin composed of 14 residues is considered as a model protein and its structure is available in protein data bank (pdb: 1j4m).

**Acknowledgements**

This work was supported by National Research Foundation of Korea under Grant No. NRF-2009-0071246 (to K.E.), and Grant No. NRF-2008-314-D00012 (to S.N.).

# Figure Captions

**Fig. 1. Quantitative descriptions on a bond rupture mechanism with respect to stiffness of loading device. a.** Normalized free energy landscape, $\bar{V}$, as a function of normalized reaction coordinate, $\bar{x}$, under the mechanical extension via loading device with respect to normalized stiffness (e.g. $\varepsilon$ = 0, 0.05, or 0.1) at a pulling rate of $\bar{\lambda}$ = 0.3. Here, we have used the linear-cubic potential to represent the intrinsic free energy landscape for a chemical bond. Moreover, the dimensionless temperature is set to $\bar{T}$ = 1. It is shown that the energy barrier is decreasing when stiffer loading device is employed to stretch a chemical bond. **b.** Dimensionless kinetic rate, $\bar{k}$, as a function of normalized stiffness of loading device, $\varepsilon$, and non-dimensional pulling rate, $\bar{\lambda}$. It is provided that kinetic rate for bond rupture is improved by using stiffer loading device. **c.** Probability distribution of bond rupture forces when a loading device with its prescribed stiffness (e.g. 0.01 N/m, 0.1 N/m, or 1 N/m) is used to stretch a chemical bond, which is depicted by linear-cubic potential with parameters set to $\Delta G_0$ = 0.24 nN·nm and $\Delta x$ = 0.3 nm, with a loading rate of $\dot{F}$ = 1 nN/s. Here, diffusion coefficient and temperature is set to $D$ = 538 nm$^2$/s and $T$ = 293 K, respectively. It is interestingly found that probability distribution of bond rupture forces depends on the stiffness of loading device. **d.** Relationship between mean bond rupture force and loading rate is obtained with respect to the stiffness of loading device. Inset shows $\Delta \bar{F}$ defined as $\Delta \bar{F} = \langle F \rangle|_{s_L} - \langle F \rangle|_{s_L = 0.01}$, which is the difference between bond rupture forces for a given force constant $s_L$ and a force constant of $s_L$ = 0.01 N/m, respectively. Inset clearly suggests that bond rupture force is dependent on the loading device. It is shown that stiffer loading device enhances the bond rupture force.

**Fig. 2. Computational simulations of a bond rupture mechanism with respect to loading device effect.** Here, we have used the Morse potential to describe the biological bond (e.g. hydrogen bond connecting phosphate group in double-stranded DNA). For more details of free energy landscape we used in the simulation, see Methods. **a.** Force exerted by a loading device (with loading rate of $\dot{F}$ = 1 nN/s) is shown with respect to stiffness of loading device. The rebinding of denatured bond is likely to be found for mechanical stretching using stiff loading device, while irreversible bond rupture is observed for mechanical extension by soft loading device. **b.** Probability distributions of bond rupture forces are governed by loading device. **c.** Mean bond rupture forces with respect to stiffness of loading device and loading rate. The softer is loading device, the smaller bond rupture force is likely to be found. This is consistent with our theoretical predictions using linear-cubic potential. This indicates that a mechanical rupture of a chemical bond is dominated by control parameters of loading device (e.g. loading rate as well as stiffness of loading device).

**Fig. 3. Computational simulations of protein unfolding mechanics with respect to loading device effect. a.** Force exerted by loading device (shown in top pannel) for mechanical stretched $\beta$-hairpin with respect to stiffness of loading device. Here, $\beta$-hairpin is mechanically stretched with loading rate of $\dot{F}$ = 3 × 10$^{-3}$ N/s. It is interestingly shown that the higher unfolding force is likely to be observed when a stiffer loading device is utilized to stretch a $\beta$-hairpin. This may be attributed to the hypothesis that a stiffer loading device is likely to induce the more dissipated energy (corresponding to higher unfolding force). In order to validate our hypothesis, we have considered the unfolding pathway for $\beta$-hairpin that is mechanically extended by loading device with different stiffness. It is shown that unfolding pathway driven by loading device with stiffness of 0.05 N/m is identical to that with stiffness of 0.01 N/m (shown in below panel). This indicates that the pathway of bond ruptures in $\beta$-hairpin is independent of loading device effect. **b.** Mean unfolding forces with respect to loading rate and stiffness of loading device. It is found that the stiffer loading device is likely to drive the higher unfolding force. This is consistent with our theoretical model of bond rupture and/or computational simulation of a chemical bond rupture. It is implied that protein unfolding mechanics governed by hydrogen bond rupture is strongly dependent on the loading devices.

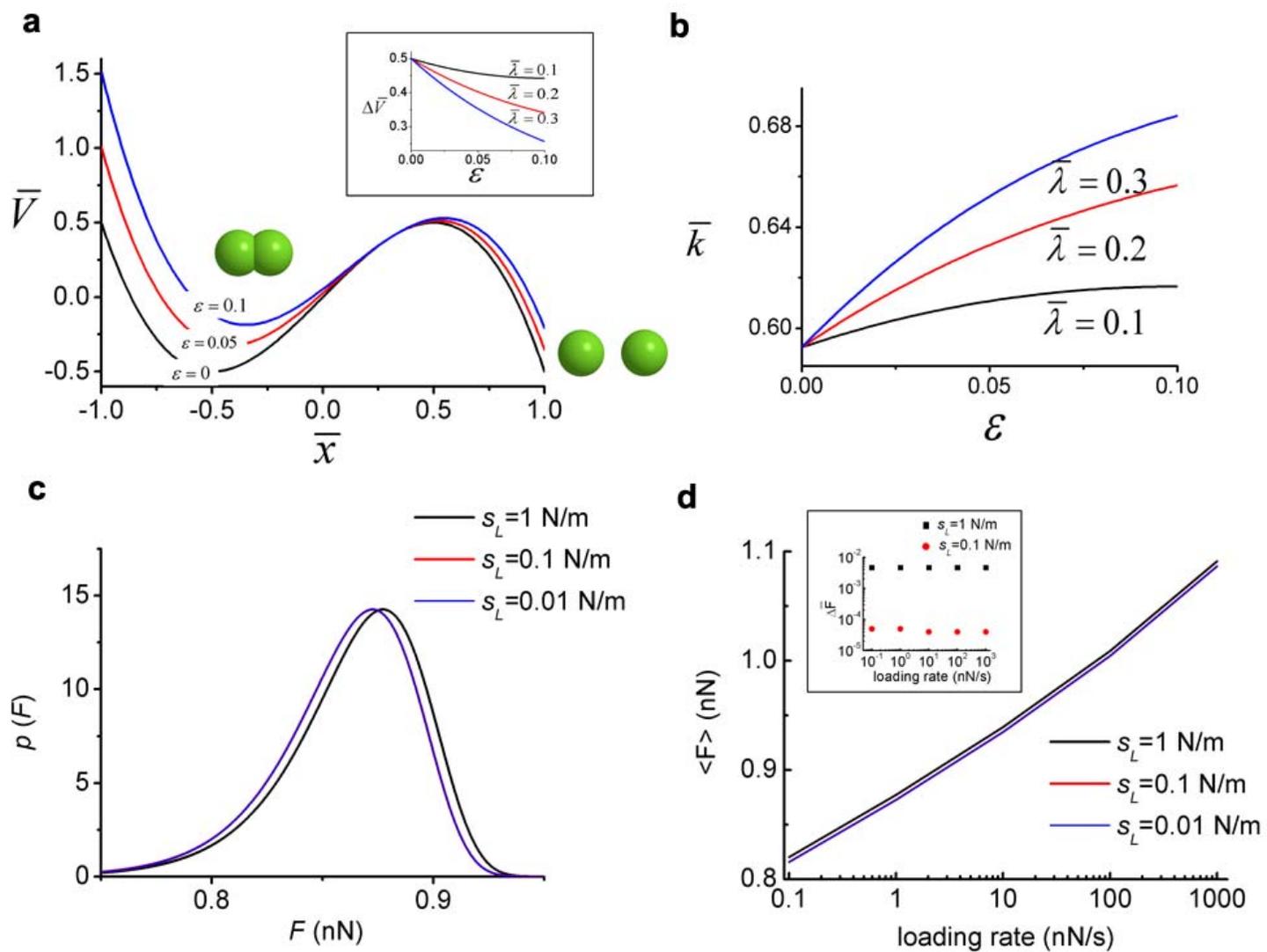

Figure 1

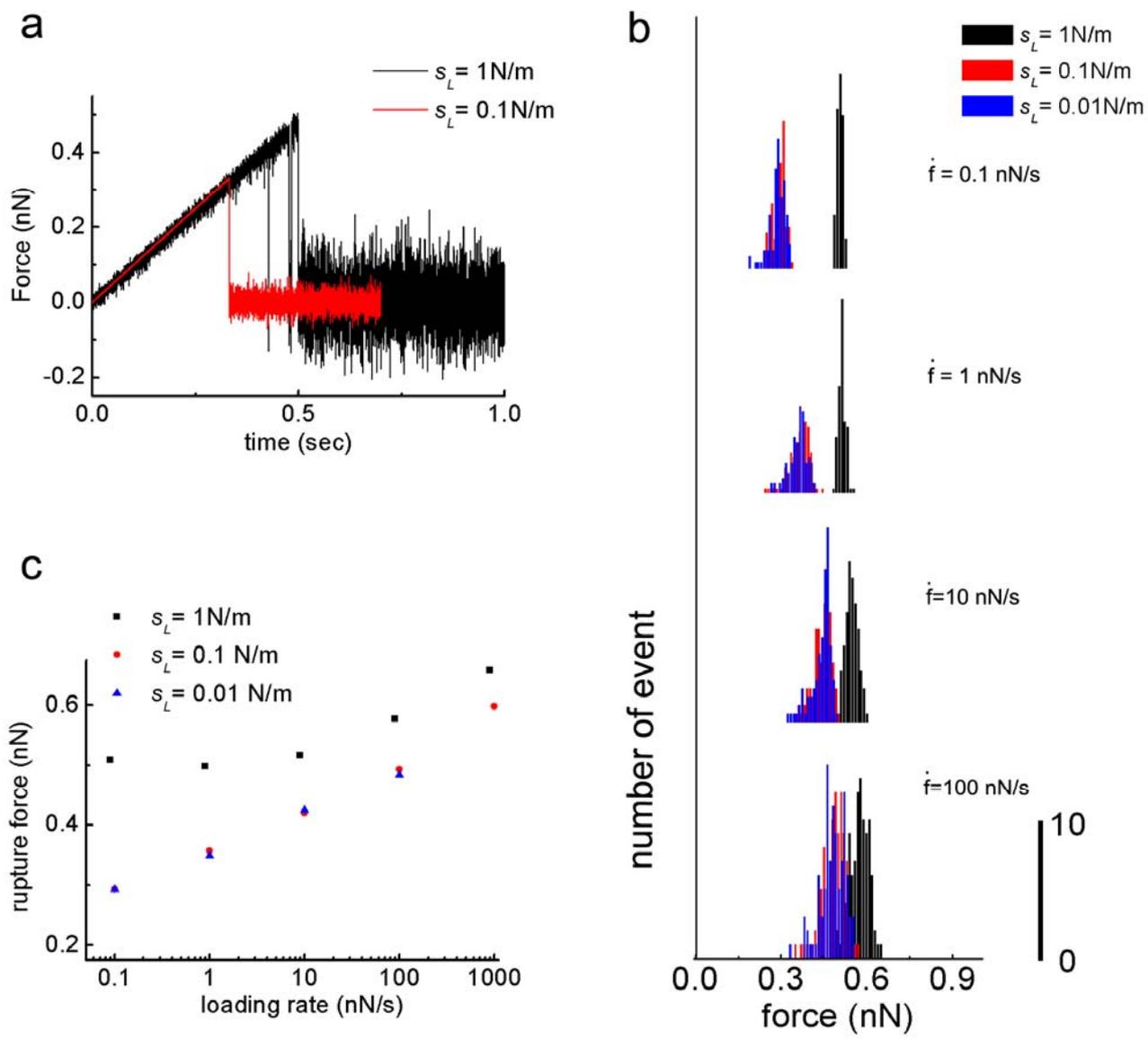

Figure 2

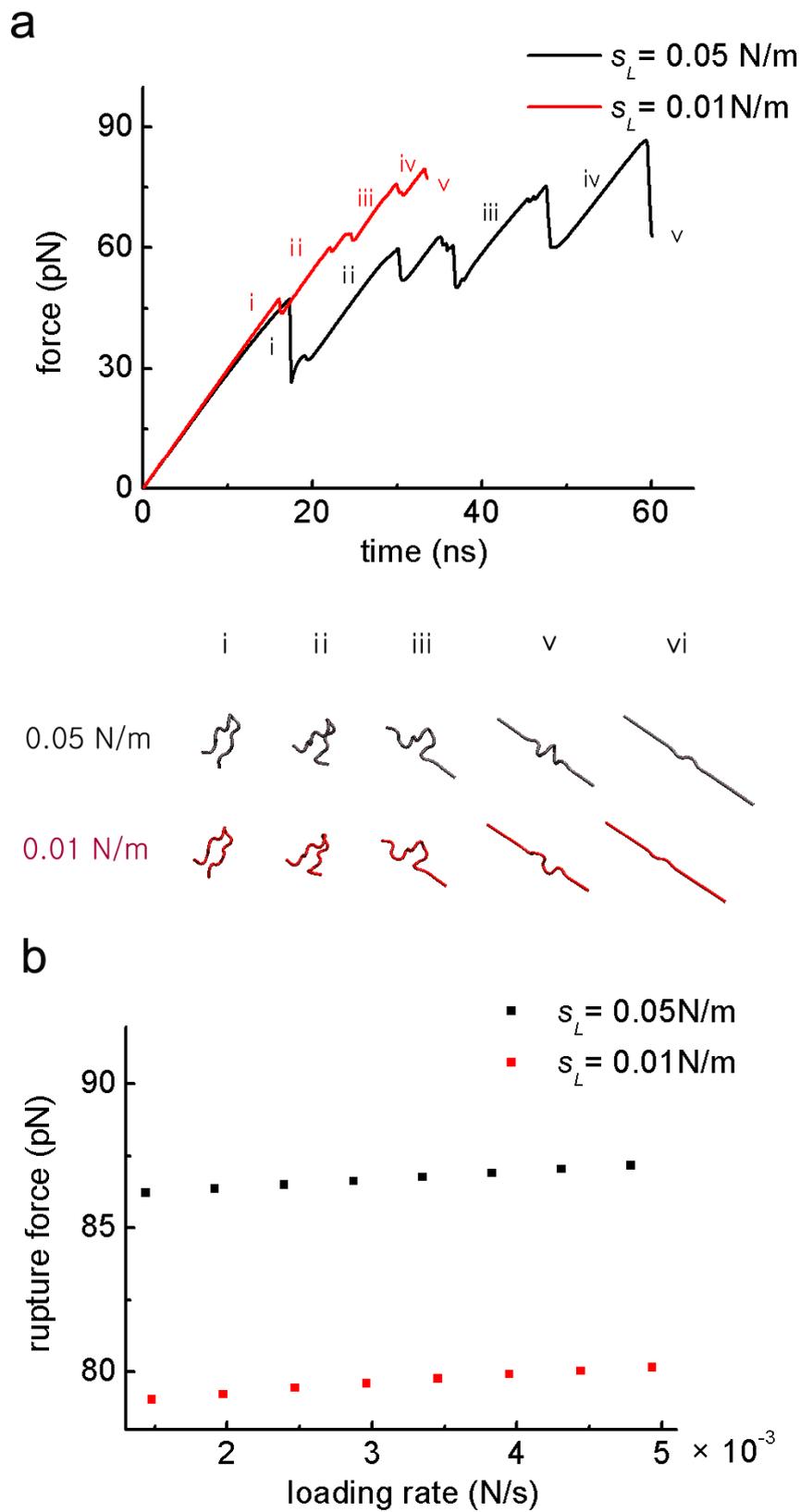

Figure 3